\documentclass[12pt]{iopart}

\usepackage{epsfig,psfig,amssymb,subfigure,float}

\begin{document}

\title[Phase separation in a boson-fermion mixture of Lithium atoms]{Phase separation in a boson-fermion mixture of Lithium atoms}

\author{Z. Akdeniz\footnote[2]{Permanent address:
Department of Physics, University of Istanbul, Istanbul, Turkey}, 
P. Vignolo\footnote[3]{To
whom correspondence should be addressed (vignolo@sns.it)}, 
A. Minguzzi, and M.~P. Tosi}
\address{NEST-INFM and Classe di Scienze, Scuola Normale Superiore,
Piazza dei Cavalieri 7, I-56126 Pisa, Italy}

\begin{abstract}
We use a semiclassical three-fluid model to analyze the conditions for
spatial phase separation  in a mixture of fermionic $^6$Li and a
(stable) Bose-Einstein 
condensate of $^7$Li atoms under cylindrical harmonic confinement, both at
zero and finite temperature. We show that with the parameters of
the Paris experiment [F. Schreck {\it et al.}, 
Phys. Rev. Lett. {\bf 87} 080403 (2001)]  an increase
of the boson-fermion scattering length by a factor five would be sufficient to
enter the phase-separated regime. We give examples of
configurations for the  density profiles in
phase separation and estimate that the transition should persist
at temperatures typical of current experiments.
For higher values of the boson-fermion coupling
we also find a new phase separation between the fermions and the
 bosonic  thermal cloud at finite temperature.
\end{abstract}
\pacs{05.30.-d, 03.75.Fi, 73.43.Nq, 67.40.Kh} 

\maketitle

\section{Introduction}      
Following the achievement of Bose-Einstein condensation in alkali gases
\cite{bec}, a great deal of experimental effort is being devoted to trap and
cool gases of fermionic isotopes of alkali atoms. The usual
evaporative cooling scheme  is ineffective,  since the Pauli
exclusion principle forbids $s$-wave collisions between spin-polarized
fermions. This problem has been circumvented by recourse to
 sympathetic cooling, which relies on $s$-wave collisions with atoms
belonging to different species,  and
several groups have thus reached in the
Fermi cloud temperatures  as low as $T=0.2\, T_F$
\cite{jila_exp,hulet_exp,paris_exp},  $T_F$ being the Fermi 
temperature. In particular, an experiment in Paris has produced a
mixture of a degenerate Fermi gas and a stable Bose-Einstein
condensate, made from Lithium isotopes \cite{paris_exp}. 

The condensate-fermion mixture is an interesting system for the study
of the effects of the interactions, since it can show spatial phase
separation  of the two components on increasing the value of the
boson-boson and boson-fermion coupling constants \cite{molmer}. This
is an example 
of a  quantum phase transition~\cite{QPT}. Several configurations for
phase-separated clouds have been proposed in~\cite{molmer_new}. A
schematic phase diagram for the trapped boson-fermion mixture at zero
temperature has been
given in~\cite{annatosi}, while the phases and the stability of a
homogeneous boson-fermion mixture have been studied in
\cite{viverit_phasediag}. In the experiments it might be possible to
investigate the phase-separated region by exploiting optically
or magnetically induced Feschbach
resonances to increase the values of the boson-boson and/or the
fermion-boson scattering length~\cite{feschbach}.

In this letter we have chosen the trap geometry and parameters typical
of the Paris experiment and  have analyzed the possible
phase-separated configurations; we also indicate how  various 
equilibrium observables are affected by  phase separation.

The paper is organized as follows: in Sec.~\ref{model} we 
present the model which we have used and discuss its limits of
applicability; Sec.~\ref{density_profiles} gives the conditions
for phase separation at zero temperature and
illustrates the  column-density profiles corresponding to different
phase-separated configurations. The results are extended to finite
temperature in  Sec.~\ref{energy}.

\section{The model}
\label{model}
We adopt a three-fluid model for a mixture of bosons and
fermions 
at finite temperature $T$~\cite{bf_model}. The model is used to evaluate
the density profile 
of the condensate ($n_c$)  within 
the Thomas-Fermi approximation and those of the fermionic ($n_f$) and
thermal bosonic 
component ($n_{nc}$)  within the Hartree-Fock approximation.

 We shall assume that the number of 
bosons in the trap
is large enough that the kinetic energy term in the Gross-Pitaevskii 
equation for the wave function $\psi({\bf r})$ of the condensate can be 
neglected~\cite{baym_pethick}.
The density profile of the condensed atoms is
\begin{equation}
n_c({\bf r})=\psi^2({\bf r})=[\mu_b-V_b^{ext}({\bf r})-2gn_{nc}({\bf r})
-fn_f({\bf r})]/g
\label{zehra1}
\end{equation}
for positive values of the function in the brackets and zero
otherwise.
Here, $g=4\pi\hbar^2a_{bb}/m_b$ and $f=2\pi\hbar^2a_{bf}/m_r$ with
$a_{bb}$ and $a_{bf}$ the boson-boson and boson-fermion $s$-wave 
scattering lengths and $m_r=m_bm_f/(m_b+m_f)$ with $m_b$ and $m_f$ 
the atomic masses; $\mu_b$ is the chemical potential of the bosons.
The system is confined by external axially symmetric
potentials
\begin{equation}
V^{ext}_{b,f}({\bf r})=m_{b,f}\omega^2_{b,f}(r^2+\lambda^2_{b,f}z^2)/2
\end{equation}
where $\omega_{b,f}$ are the frequencies and $\lambda_{b,f}$ the
anisotropies of the traps.

As already proposed in early work on the confined Bose
fluid~\cite{anna_tesi},
we treat both the thermal bosons and the fermions as ideal gases in effective potentials $V^{eff}_{b,f}({\bf r})$ involving the relevant interactions.
We write
\begin{equation}
V^{eff}_{b}({\bf r})=V^{ext}_{b}({\bf r})+2gn_c({\bf r})+2gn_{nc}({\bf r})
+fn_f({\bf r})
\label{zehra2}
\end{equation}
and
\begin{equation}
V^{eff}_{f}({\bf r})=V^{ext}_{f}({\bf r})+fn_c({\bf r})+fn_{nc}({\bf r}).
\label{zehra3}
\end{equation}
We are taking the fermionic component as a dilute, spin-polarized Fermi gas: 
the fermion-fermion interactions are then associated at leading order with 
$p$-wave scattering and are  negligible at the temperatures of 
present interest~\cite{pwave}. Here and in Eq.~(\ref{zehra1}), the factors 2
arise from exchange. 

We may then evaluate the thermal averages by means of standard
Bose-Einstein and  
Fermi-Dirac distributions, taking the thermal bosons to be in thermal 
equilibrium with the condensate at the same chemical potential and the 
fermions at chemical
potential $\mu_f$. In this semiclassical approximation  
the  densities of the  cold-atom clouds are
\begin{equation}
n_{nc,f}({\bf r})=\int\frac{d^3p}{(2\pi\hbar)^3}\left\{\exp\left[\left(
\frac{p^2}{2m_{b,f}}
+V^{eff}_{b,f}({\bf r})-\mu_{b,f}\right)/k_BT\right]\mp 1\right\}^{-1}.
\label{zehra4}
\end{equation}
The chemical potentials are determined from the 
total numbers of bosons and fermions.
%\begin{equation}
%N_b=\int d^3r[n_c({\bf r})+n_{nc}({\bf r})]
%\label{zehra5}
%\end{equation}
%and
%\begin{equation}
%N_f=\int d^3 r\,n_f({\bf r}).
%\label{zehra6}
%\end{equation}
%These equations complete the self-consistent closure of the model.

The above model is valid for bosons when the diluteness condition $n_c
a_{bb}^3 \ll 1$ holds. For the Fermi cloud in the mixed
 regime the condition $k_f a_{bf} \ll 1$ with $k_f=(48
N_F)^{1/6}/a_{ho}$ should
hold, but 
 this is not a constraint in the regime of phase separation where
the boson-fermion interaction energy drops rapidly (see below).

\section{Conditions for phase separation}

\label{density_profiles}
\begin{figure}
\begin{center}
\epsfig{file=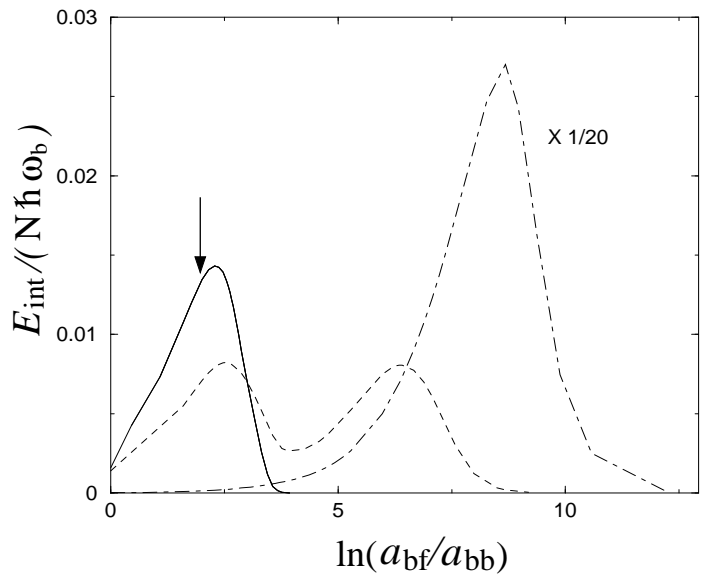,width=0.6\linewidth}
\end{center}
\caption{Boson-fermion interaction energy (in units of $N\hbar\omega_b$
with $N=20000$ being the total number of particles), as a
function of $\ln(a_{bf}/a_{bb})$. The curves are obtained for 
$a_{bb}=5.1\,a_0$ and equal numbers of bosons and fermions
at various values of the temperature: $T=0$ (solid line),
$T=0.2\,T_F$ (dashed line) and $T=0.6\,T_F$ (dotted-dashed line, on a
scale reduced by a factor 20).
The arrow indicates the location of the Paris experiment.}
\label{phase_sep}
\end{figure}

Phase separation is complete when the overlap between the
bosonic and the fermionic density profiles vanishes.
The  transition to this regime depends on 
the confinement and
on the numbers, masses and interaction strengths
of the trapped fermions and bosons.
In the case of equal numbers of bosons and
fermions 
and equal trapping frequencies
the condition of phase separation
can be written as
\begin{equation}
k_fa_{bb}>\frac{3\pi}{4}\left(\frac{a_{bb}}{a_{bf}}\right)^2
\label{condition}
\end{equation}
at zero temperature~\cite{annatosi,viverit_phasediag} .

Let us focus on the geometry and on  typical values of the numbers of atoms
in the Paris experiment~\cite{paris_exp} for the $^7$Li-$^6$Li mixture:
$\omega_b=2\pi\times4000$~Hz, $\omega_f=2\pi\times3520$~Hz, 
$\lambda_b\simeq\lambda_f\simeq 1/60$, 
and $N_b \simeq N_f \simeq 10000$.
Since the two trap frequencies
are of the same order of magnitude, we can use Eq.~(\ref{condition})
to estimate at which values of the scattering lengths the 
phase transition occurs. 
For the above  experimental parameters the condition~(\ref{condition})
becomes 
$\xi\equiv a_{bf}^2/a_{bb}>7000\,a_0$.
Using the predicted 
  values  for the scattering lengths of the $^7$Li-$^6$Li
mixture, that is  $a_{bb}=5.1\, a_0$  and $a_{bf}=38 \, a_0$ \cite{paris_exp},
we have $\xi\simeq 283\,a_0$ and the gases are not as yet in  the 
phase-separated regime. 
However, in a Feschbach-resonance experiment the scattering lengths could be
increased. There is therefore
the possibility to fulfill
Eq.~(\ref{condition}) and to enter the phase-separated regime.

Since the positions of the Feschbach resonances for $a_{bf}$ and 
$a_{bb}$ would not
 necessarily coincide, in order
 to explore the phase diagram for the boson-fermion mixture we have
 first of all kept $a_{bb}$ fixed at its non-resonant value and 
 changed $a_{bf}$. In this case, in order to reach phase separation
it is sufficient to increase the value of the boson-fermion scattering length
by a factor five, namely $a_{bf}> \tilde{a}_{bf}\simeq 200\,a_0$.

To check this estimate we have evaluated the boson-fermion
interaction 
energy 
\begin{equation}
E_{int}=f \int d^3 r\,[n_c({\bf r})+n_{nc}({\bf r})]n_f({\bf r})\;,
\end{equation}
using the model described in Sec.~\ref{model}.
The behaviour of this quantity at $T=0$
as a  function of $a_{bf}/a_{bb}$, with  $a_{bb}=5.1 a_0$, is shown in
Fig.~\ref{phase_sep} (solid line). 
Because of finite size effects the transition to the
phase-separated regime shows a large crossover region and the
interaction energy goes smoothly to zero  when separation is complete 
\cite{noticina}.
%, the phase transition does not occur with
%a discountinous derivative in the interaction energy, which in fact
%goes to zero with a rather slight slope. 

\begin{figure}[H]
\vspace{-1cm}
\centerline{
\epsfig{file=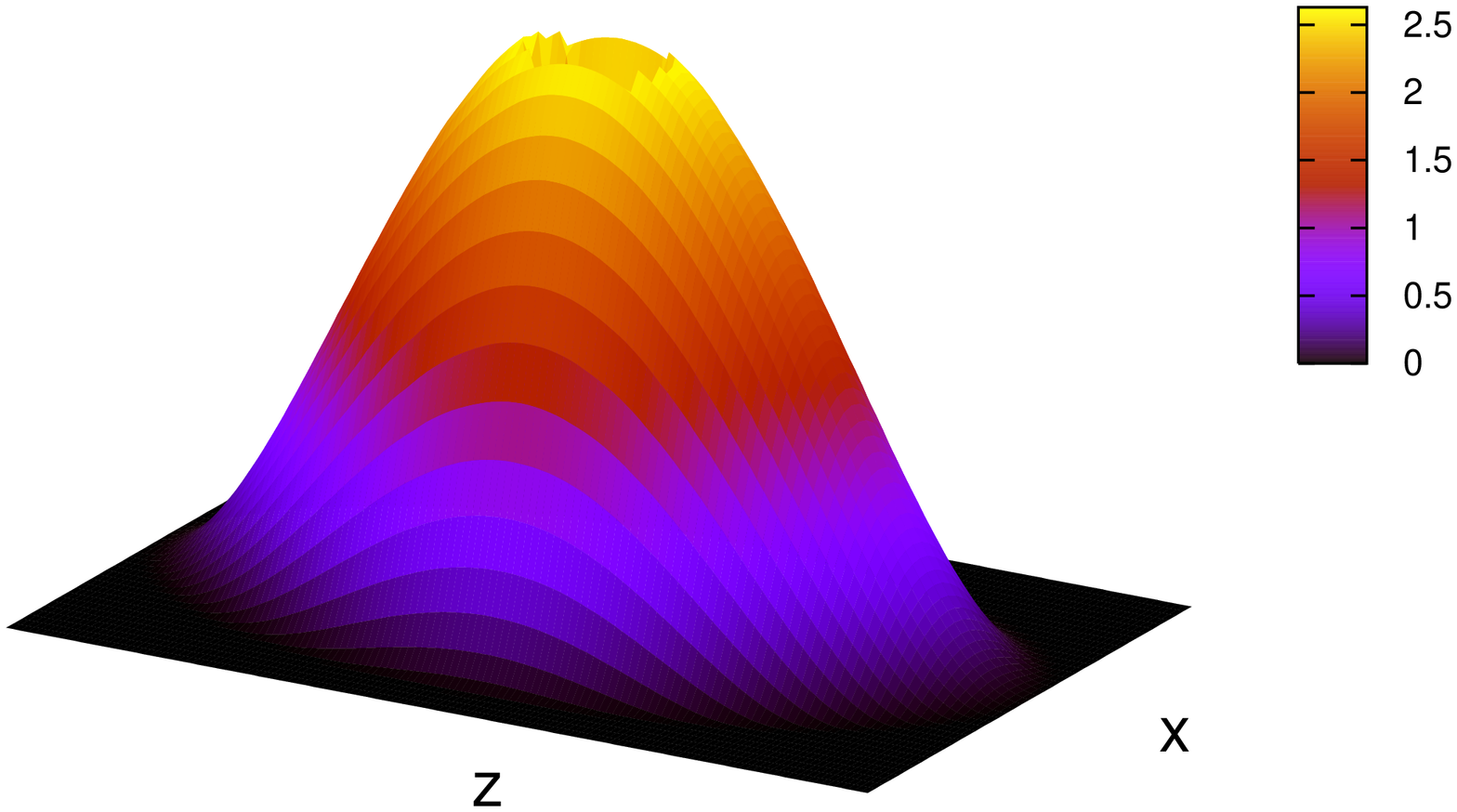,width=0.40\linewidth,angle=270}
\hspace{-2cm}
\epsfig{file=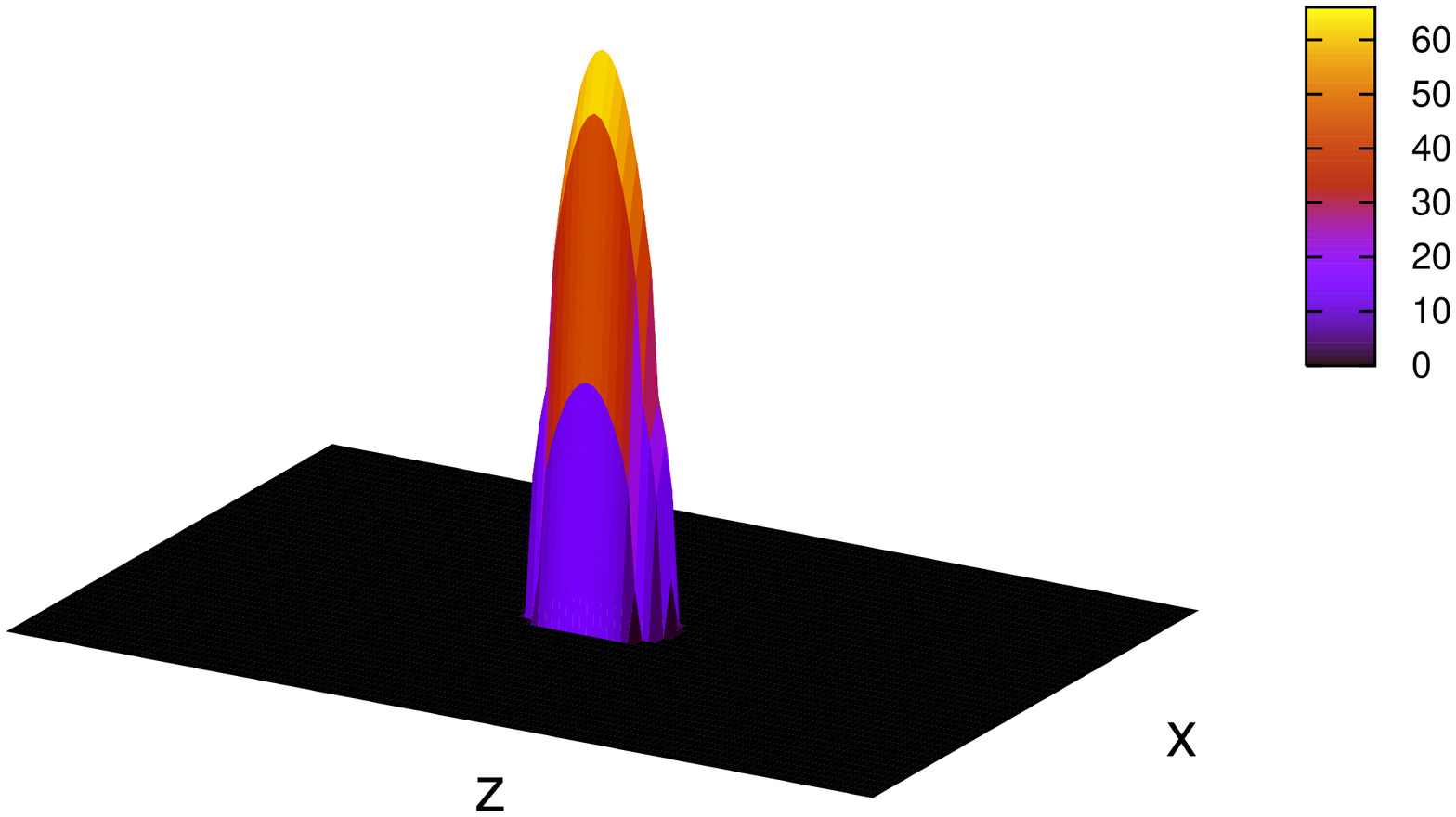,width=0.40\linewidth,angle=270}}
\vspace{-2cm}
\centerline{
\epsfig{file=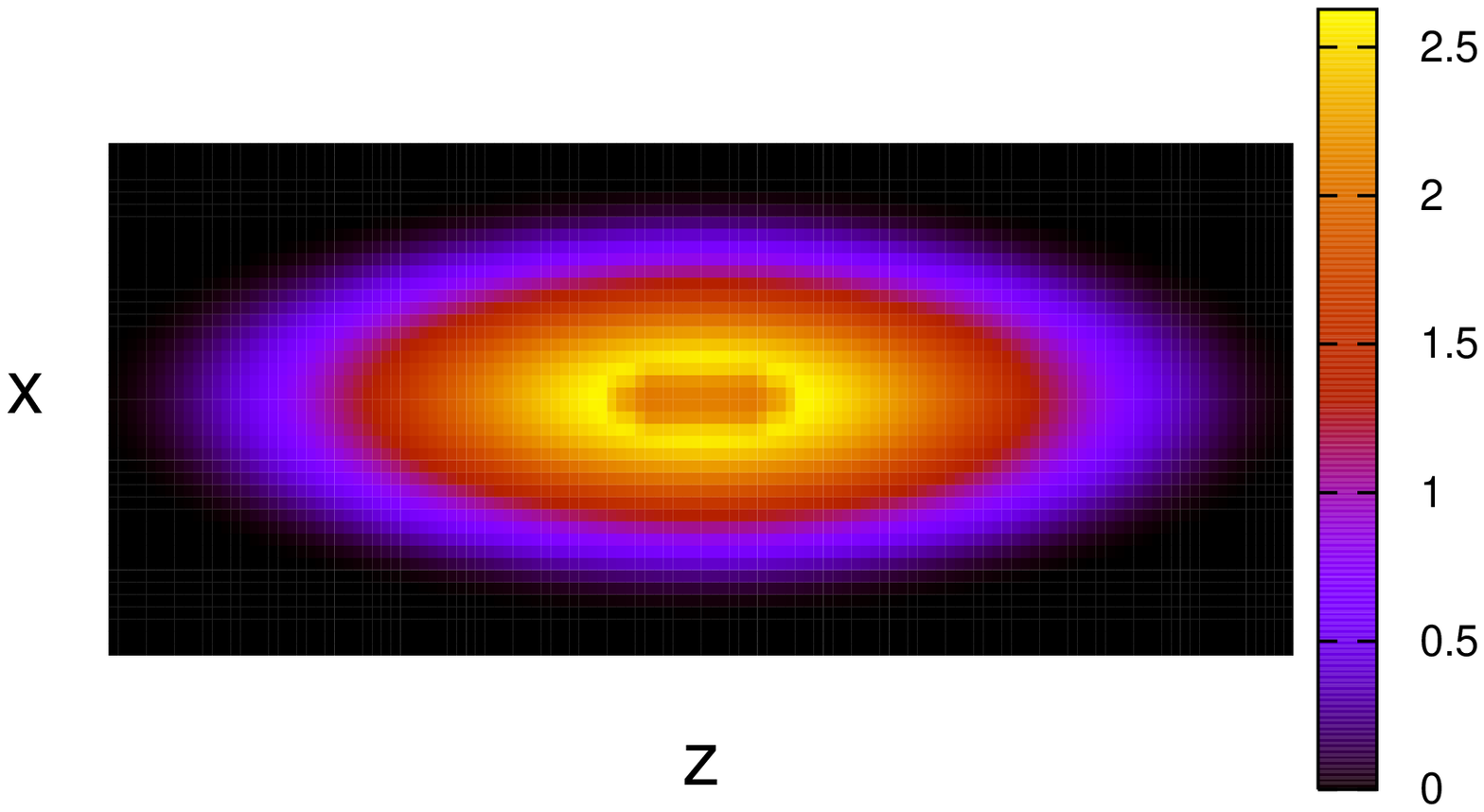,height=0.55\linewidth,angle=270}
\epsfig{file=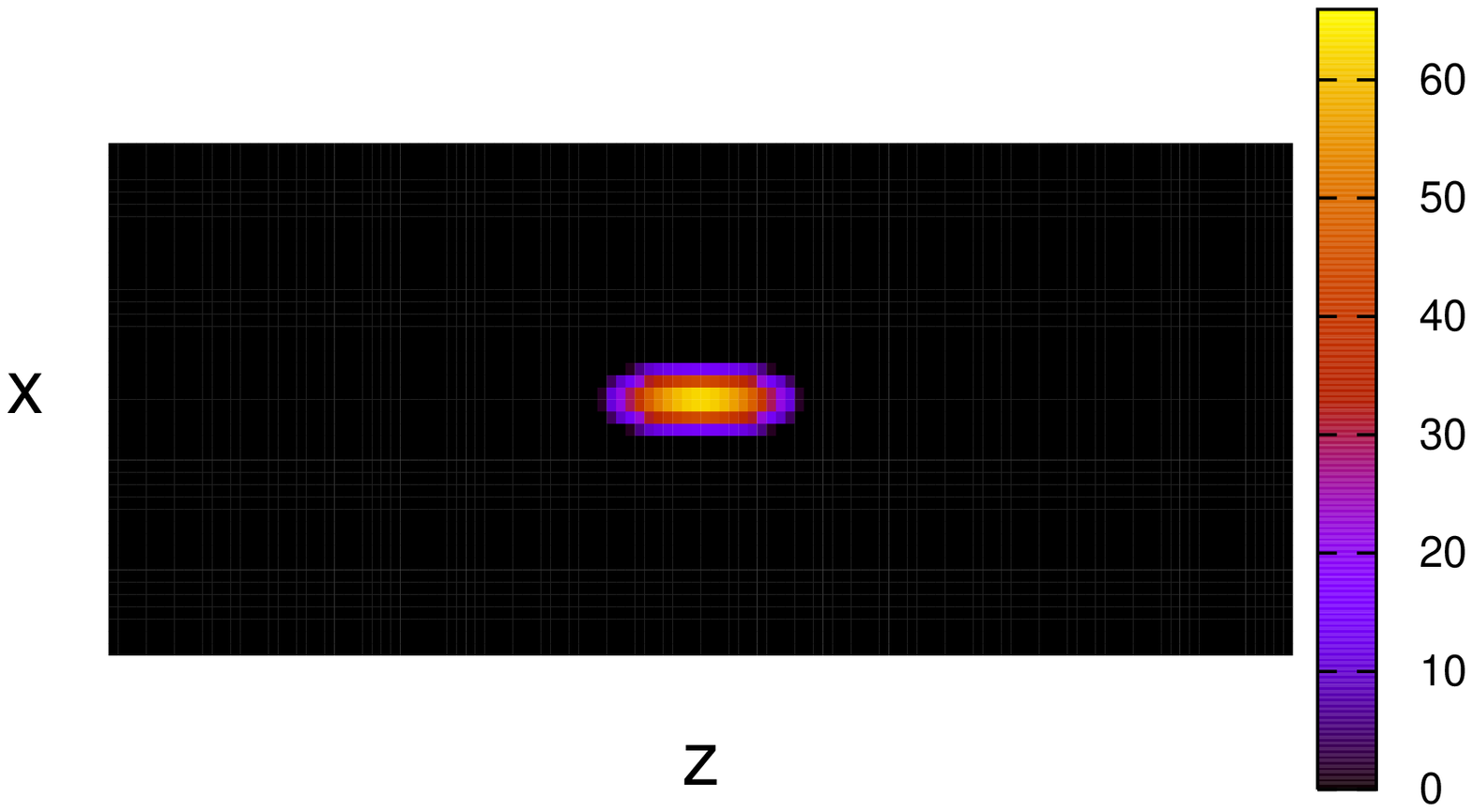,height=0.55\linewidth,angle=270}}
\vspace{-1cm}
\caption{The symmetric configuration at the threshold of the 
phase-separation regime, obtained for $a_{bb}=5.1\,a_0$ and
$a_{bf}=200\,a_0$. 
Top panels: column densities in the $\{x,z\}$ plane of the fermionic
cloud (left)   
and of the condensate (right); bottom panels: topview images of the
same profiles.
The size of the figures in the $\{x,z\}$ plane is 10$\mu$m $\times$
665$\mu$m.} 
\label{fig200}
\end{figure}
\begin{figure}[H]
\vspace{-1cm}
\centerline{
\epsfig{file=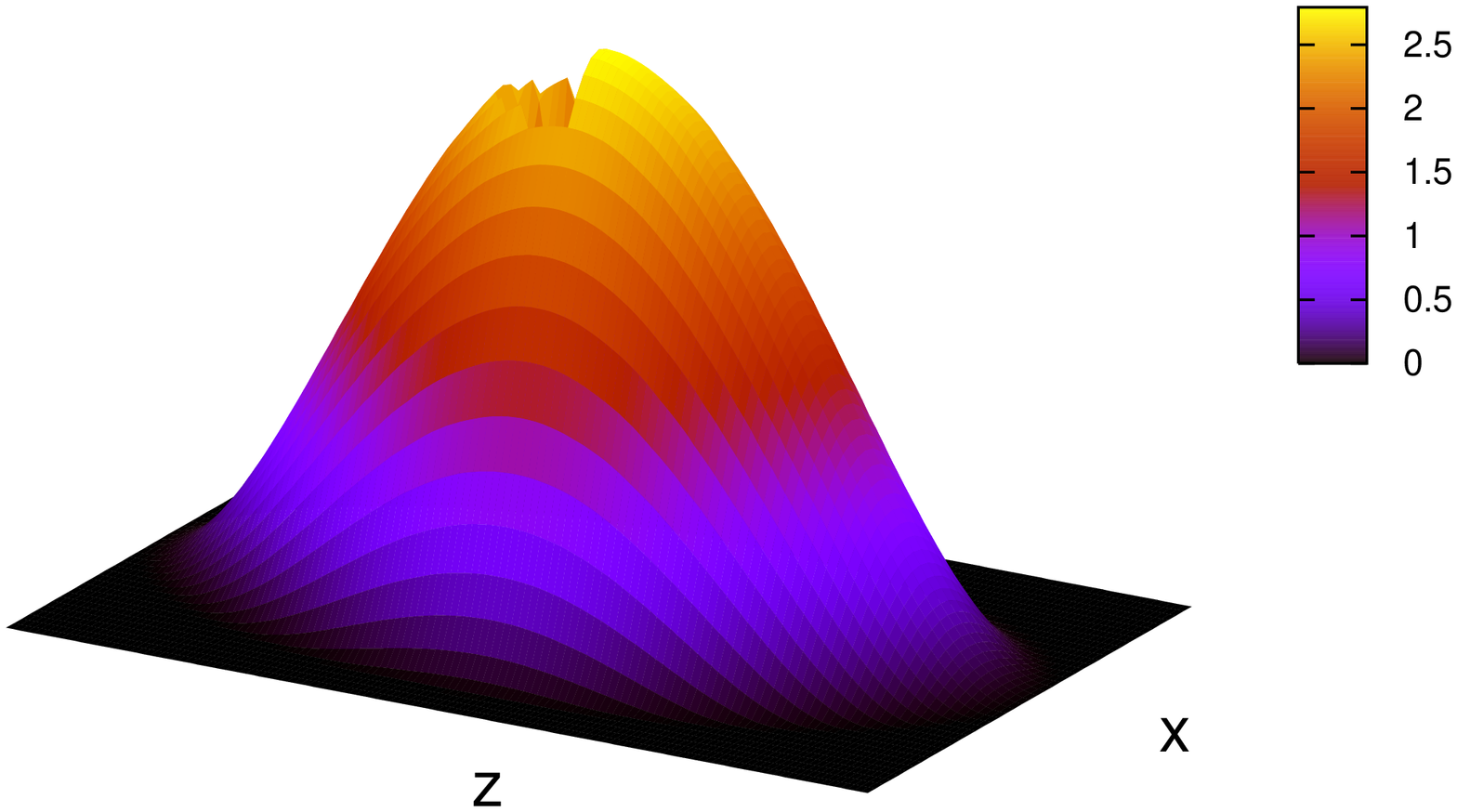,width=0.40\linewidth,angle=270}
\hspace{-2cm}
\epsfig{file=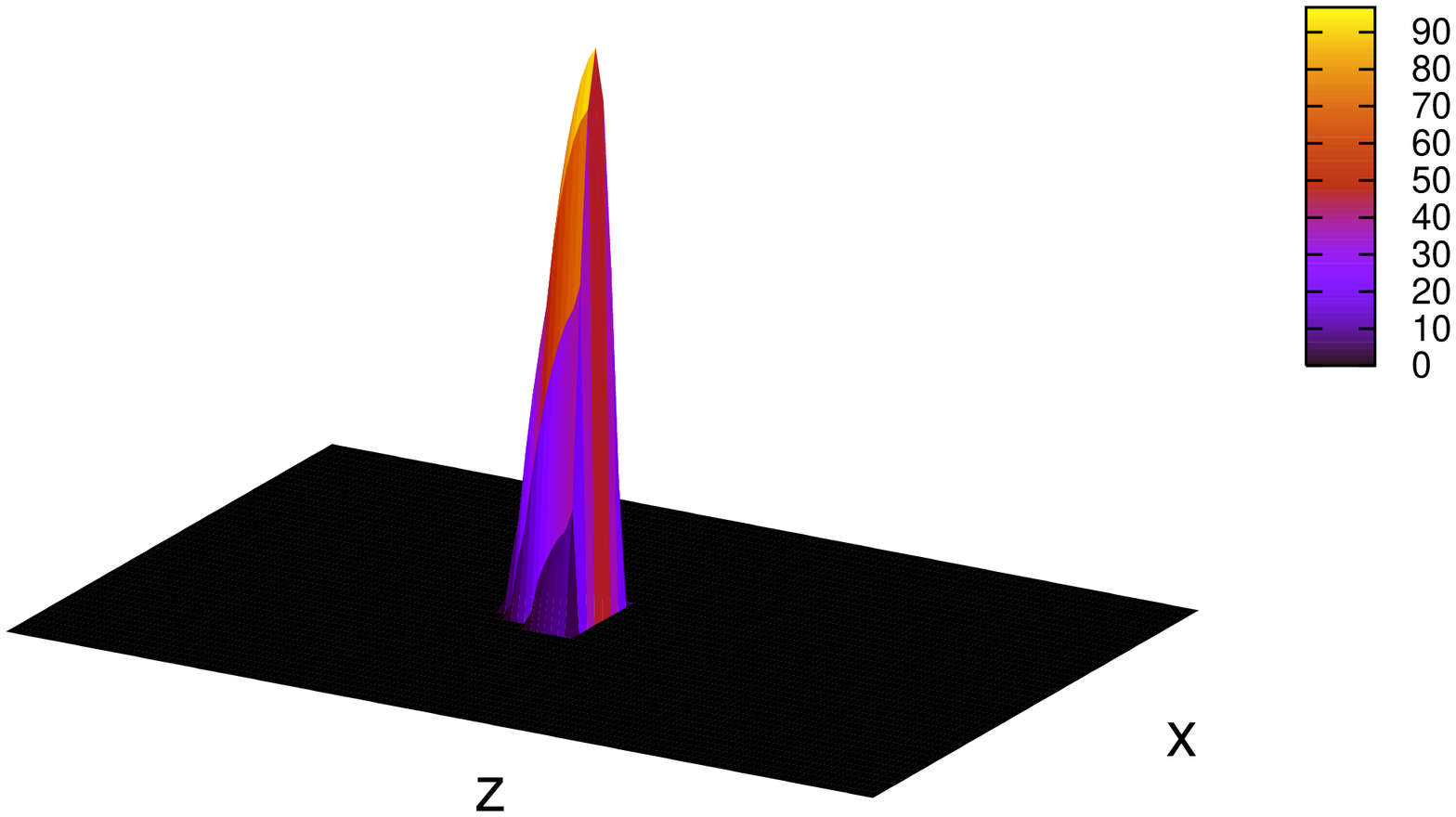,width=0.40\linewidth,angle=270}}
\vspace{-2cm}
\centerline{
\epsfig{file=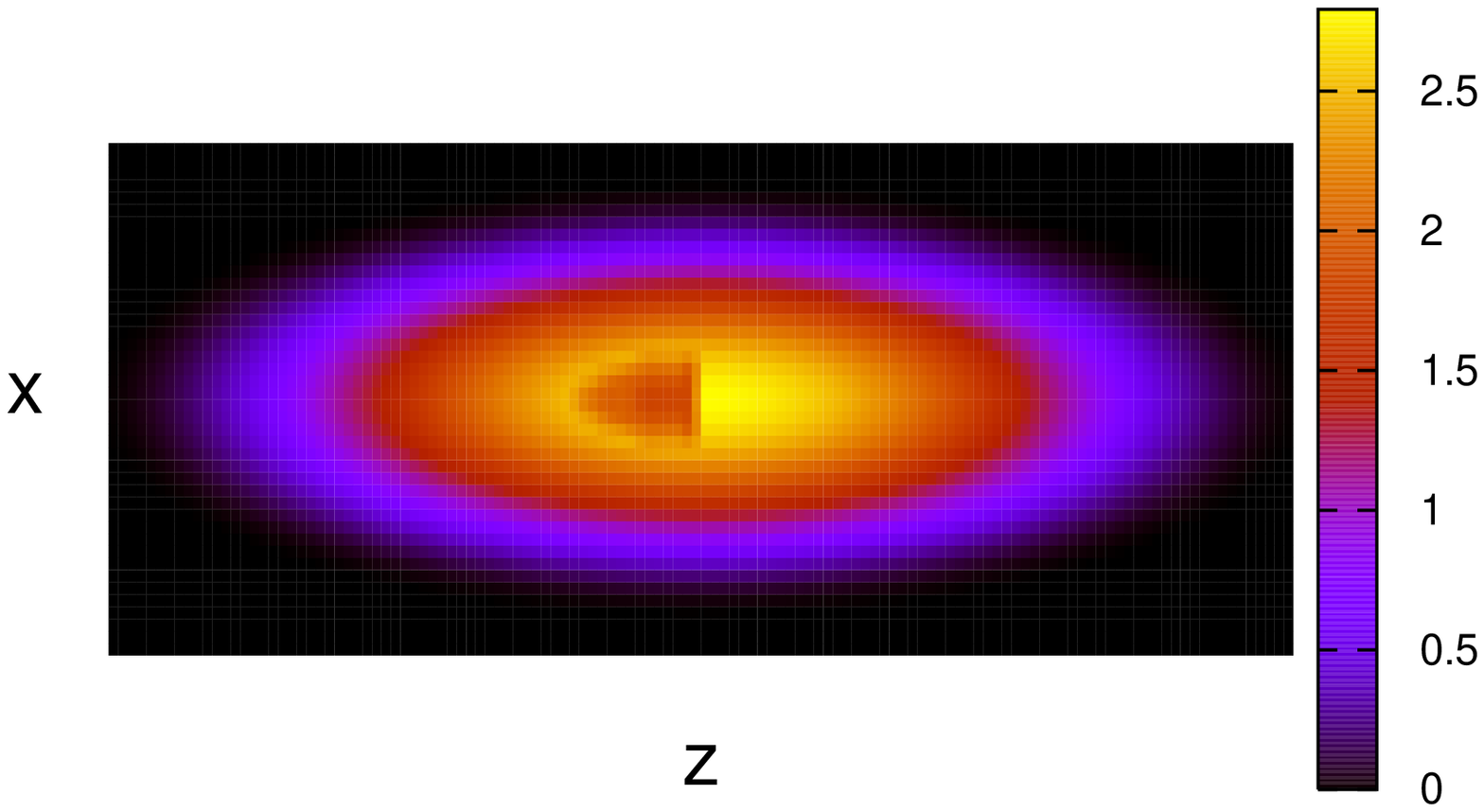,height=0.55\linewidth,angle=270}
\epsfig{file=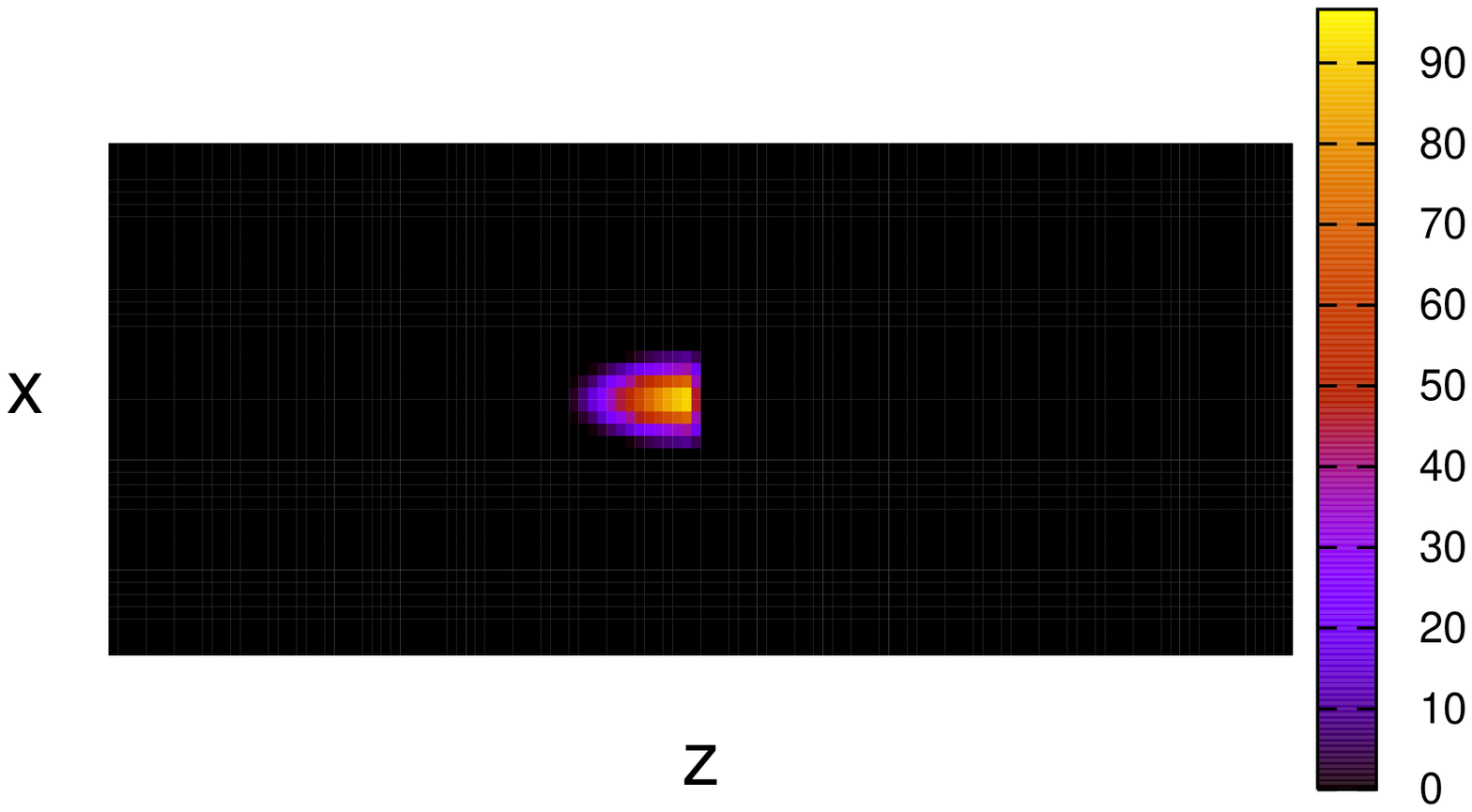,height=0.55\linewidth,angle=270}}
\vspace{-1cm}
\caption{The fully asymmetric configuration for 
$a_{bb}=5.1\,a_0$ and $a_{bf}=1000\,a_0$. Top panels:
column densities in the $\{x,z\}$ plane of the fermionic cloud (on the left)
and of the condensate (on the right); bottom panels: topview images
of the same profiles.
The size of the figures in the $\{x,z\}$ plane is 10$\mu$m $\times$
665$\mu$m.}
\label{fig1000}
\end{figure}

For the value $a_{bf}=38 a_0$ 
corresponding to the current experiments (marked by the arrow in
Fig.~\ref{phase_sep}), the 
boson-fermion 
interaction energy is close to its maximum value. We find
that the overlap between the two clouds 
vanishes if $a_{bf}> 200 a_0$, in accord with the estimate
given just above.

To make close  contact with experiments we have calculated
the column densities of the fermionic and condensate vapours
in the phase-separated regime for various values of $a_{bf}$.
At the threshold value we have found only one possible configuration,
with a central hole in the fermionic cloud which is  occupied by the condensed
bosons (see Fig.~\ref{fig200}).  For this configuration
 there is only a quantitative
difference
in the value of the contrast at the center of the fermionic
cloud as compared to the mixed state. 

The regime of phase separation is univocally  characterized by the
presence of other configurations
at  higher values of $a_{bf}$.
Starting from
different initial conditions, for $1000\,a_0 <a_{bf}<2000\,a_0$ we
have found, in addition to a  
configuration  
which is similar to that in Fig.~\ref{fig200}, another with a different
ellipticity of the hole and a third  one which is asymmetric in the
axial direction (see Fig.~\ref{fig1000}).
For the values of the parameters  that we have used  the lowest energy
configuration 
is the first (symmetric) one. However, also the other metastable states
are close in energy and 
could perhaps be realized in the laboratory  by applying an adiabatic 
perturbation having  the symmetry of the final state.
The asymmetric state could  easily be identified in the
column-density pictures.

 Other  interesting configurations would be observed in an experiment
in which both  scattering lengths could be 
changed. 
On  increasing
$a_{bb}$  the size of the condensate increases towards
that of the fermion cloud 
  and the hole in the latter widens, until  a  situation
is reached in which
the fermions 
are pushed out in the radial direction and make a ring around
a cigar of bosons. Another peculiar shape is a ``double
sandwich'' of bosons, in which two slices of condensate
are separated by a central slice of fermions and surrounded by a
fermionic shell.
As a last   remark,  the energy of all these configurations depends
on the values of the scattering lengths: it may thus happen that
the most stable state is the asymmetric one or has
other exotic shapes~\cite{noipoi}.

\section{Effects of temperature}
\label{energy}

\begin{figure}
\begin{center}
\epsfig{file=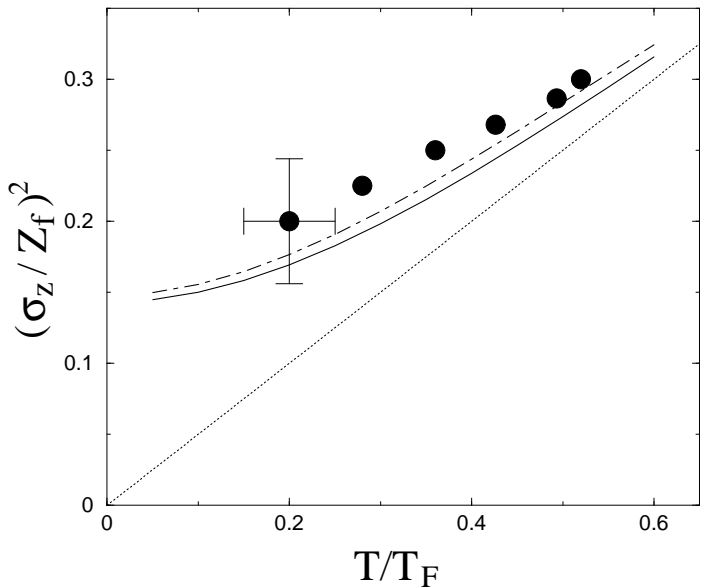,width=0.6\linewidth}
\end{center}
\caption{Variance of the axial distribution of a cloud of 
10000 atoms of $^6$Li taken at $x=0$ (in units of
$Z_f^2=2k_BT_F/m_f\omega_f^2$) against
$T/T_F$ with $T_F=(6N_f)^{1/3}\hbar\omega_f/k_B$.
The experimental data (filled circles;
from Ref~\protect\cite{paris_exp} and private communication)
 are compared with the behaviour of the 
classical gas (dotted line), of the
ideal Fermi gas (full line),
and of the fermionic cloud interacting with its bosonic 
partner ($N_b=20000$) 
using the  values 
$a_{bb}=5.1\,a_0$ and $a_{bf}=200\,a_0$ (dotted-dashed line).}
\label{temp}
\end{figure}

At low temperature ($T\sim 0.2\,T_F$) the density profiles for the various
configurations are essentially the same as those found at $T=0$,
with the addition of a small bosonic thermal cloud.
For values of the boson-fermion scattering length around 
the critical value $ \tilde{a}_{bf}$
the phase separation involves only  the condensate and the
fermionic cloud:  
the total boson-fermion interaction energy $E_{int}$
has a  minimum but does not vanish (see 
the dashed line in Fig.~\ref{phase_sep}), since the fermionic and the
bosonic thermal clouds still overlap. However, with a further
increase in the scattering length, 
up to values of order $10^5\, a_0$, the model yields 
full phase separation
between fermions and bosons.

At higher temperature ($T\sim 0.6\,T_F$)  we do not find phase separation
between the fermionic cloud and the condensate (see
the dotted-dashed line in
Fig.~\ref{phase_sep}). 
Separation between the total bosonic gas and the fermions 
is instead found  in this model at $a_{bf}\sim 10^5\,a_0$. 
In this case a symmetric configuration is generated, in which
 the fermions 
are pushed outside the whole boson gas.
The size of the hole in the fermionic cloud
is determined by the width of the bosonic
thermal cloud, making the finite-temperature 
transition much more evident in the column-density images  than  
at $T=0$~\cite{noipoi}.

Another quantity which can be measured as a function of 
temperature is the width of the fermionic density profile in a given
direction, {\it e.g.} in the $z$ direction (see Fig.~\ref{temp}). While for
the values of the scattering lengths in the Lithium experiment the width is
predicted to be almost the same as that of the non-interacting Fermi
gas \cite{commento}, larger 
increases in width are found in some  phase-separated
configurations. This  observable is particularly
sensitive to the configurations which involve a 
large displacement of the fermions, such as the ``double boson sandwich''
\cite{noipoi}.
  
In conclusion, we have examined the main observable features of spatial
phase separation in gaseous mixtures of Li isotopes, with specific
focus on the setup of the Paris experiments. A full account of our
results will be given elsewhere~\cite{noipoi}.

We acknowledge support from MIUR under PRIN2000 and by INFM under
PRA2001.

%\begin{figure}
%\begin{center}
%\epsfig{file=abf6000_dir.ps,width=0.5\linewidth,angle=270}
%\end{center}
%\centerline{
%\epsfig{file=fermi_xz_abb600abf6000dir.ps,width=0.48\linewidth,angle=270}
%\hspace{-3cm}
%\epsfig{file=bose_xz_abb600abf6000dir.ps,width=0.48\linewidth,angle=270}}
%\caption{$a_{bb}=600\,a_0$, $a_{bf}=6000 a_0$ sandwich configuration}
%\end{figure}


\begin{thebibliography}{100}
\bibitem{noticina}
In an infinite system this quantum phase transition is
instead characterized by a discontinuity in the interaction energy as
a function of $a_{bf}/a_{bb}$.
\bibitem{commento}
Within the error bars, the model is in agreement with
the experiments for the magnitudes of the widths of the fermionic and
bosonic clouds.
\end{thebibliography}
\end{document}